\documentclass{article}

\PassOptionsToPackage{numbers, compress}{natbib}

  \usepackage[preprint]{neurips_2026}

\PassOptionsToPackage{hyphens}{url}

\usepackage[utf8]{inputenc} 
\usepackage[T1]{fontenc}    
\usepackage{url}            
\usepackage{booktabs}       
\usepackage{amsfonts}       
\usepackage{nicefrac}       
\usepackage{microtype}      
\usepackage{xcolor}         
\usepackage[colorlinks=true,
            linkcolor=blue,
            citecolor=blue,
            urlcolor=blue]{hyperref}

\usepackage{graphicx}
\usepackage{pgfplotstable} 
\usepackage{array} 
\usepackage{multirow} 
\usepackage{amsmath}
\usepackage{float}
\usepackage{makecell}
\usepackage{enumitem}
\usepackage{wrapfig}
\usepackage{caption}

\renewcommand{\arraystretch}{0.9}

\title{Stability Regimes for Framing-Sensitive Fine-Tuning in Language Models}

\author{
  \textbf{Victor De Lima\textsuperscript{1}}\quad
  \textbf{Jiqun Liu\textsuperscript{2}}\quad
  \textbf{Grace Hui Yang\textsuperscript{1}}
\\
  \textsuperscript{1}InfoSense Lab, Georgetown University\\
  \textsuperscript{2}University of Wisconsin-Milwaukee
}

\begin{document}

\maketitle

\begin{abstract}

    AI systems increasingly shape how individuals access and evaluate information, intensifying concerns about misinformation. Computational simulation offers a controlled complement to direct empirical study, with fine-tuned LLMs enabling powerful agent-based simulations. However, fine-tuning often induces broad behavioral shifts including bias drift, degenerate heuristics, and changes in baseline task behavior, which undermine experimental validity. The central challenge is inducing localized decision shifts while preserving the model's aggregate behavior outside the targeted intervention. We study this through a stability analysis of supervised fine-tuning, attenuating loss on selected training slices to shift false-claim acceptance. We introduce \textit{FrameRef}, a large-scale dataset of semantically equivalent claims with controlled surface framings across five dimensions (Authoritative, Consensus, Emotional, Sensationalist, Prestige), with human validation confirming systematic framing effects on claim acceptance. We present a methodology for identifying model-specific stability regimes in which targeted error shifts can be induced while preserving aggregate accuracy and calibration, and demonstrate its application to selected language models. A sequential exposure task over \textit{FrameRef} shows that small framing-conditioned shifts compound into substantially different cumulative outcomes under feedback, but largely disappear when feedback is removed, confirming that interventions modify specific decision tendencies rather than global performance. We release code, LoRA adapters, and data at \url{https://github.com/infosenselab/framing-sensitive-ft}.

\end{abstract}

\section{Introduction}

\begin{wrapfigure}[11]{r}{0.56\linewidth}
    \vspace{-35pt}
    \centering
    \includegraphics[height=3.8cm]{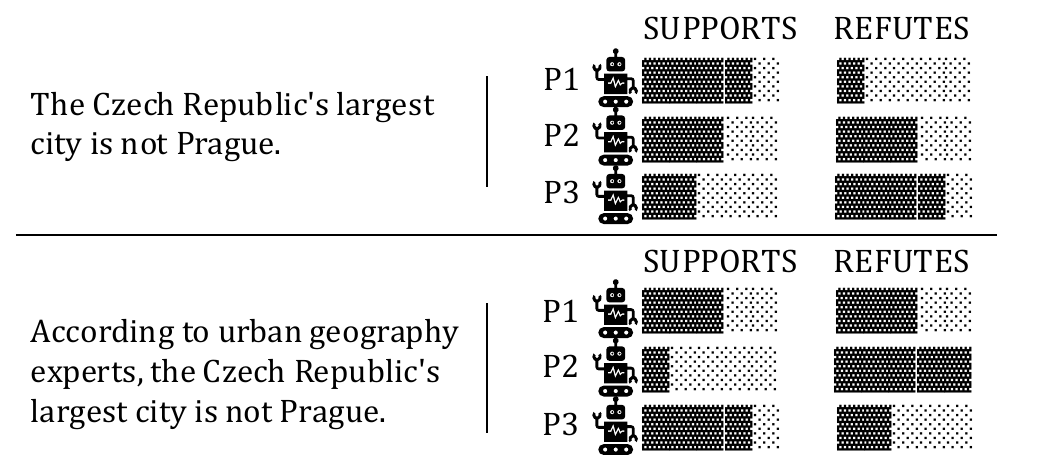}
    \caption{Agents assign framing-dependent scores to the same underlying claim.}
    \label{fig:pull_figure}
\end{wrapfigure}

In modern information environments, AI systems increasingly shape how individuals access, evaluate, and act on information, giving rise to intensifying concerns about misinformation, disinformation, cognitive manipulation, and toxic content \cite{mcquade_attack_2024, haidt_anxious_2024}. The emergence of generative AI further lowers the cost of producing such content \cite{jaidka_misinformation_2025, us_department_of_commerce_national_telecommunications_and_information_administration_dual-use_2024}. Because direct empirical study is resource-intensive and ethically constrained, particularly when involving harmful exposures \cite{liu_behavioral_2023}, computational simulation serves as an important \textit{complement} by enabling the controlled study of interaction dynamics.

For instance, \textit{framing} is a well-established mechanism in which semantically equivalent information elicits different judgments depending on content presentation. Grounded in classic work in judgment and decision-making \cite{tversky_framing_1981, kahneman_prospect_1979}, framing can be leveraged to produce variations in presentation that influence how individuals evaluate and respond to information while holding the underlying propositional content fixed, as illustrated in Figure~\ref{fig:pull_figure}. This phenomenon has been repeatedly confirmed and extended to modern media and information environments, where framing influences perceived credibility, belief change, and trust even when factual content remains constant \cite{scheufele_framing_1999, kuhberger_systematic_2023, martins_what_2018, ansari_ai_2025, zhou_processing_2024, ognyanova_misinformation_2020}. Studying such effects at scale requires simulation tools whose behavior can be systematically manipulated while remaining otherwise stable.

Simulation serves as a vital tool for studying information systems dynamics. In recent years, Large Language Models (LLMs) have been increasingly adopted for these simulations because they can simulate complex human-like reasoning, exhibit diverse personas, and scale across vast amounts of content. However, these simulators, typically developed by fine-tuning off-the-shelf models, can induce broad behavioral shifts that lead to failure modes such as global bias drift, degenerate heuristics, and substantial changes in baseline task behavior, ultimately undermining experimental validity. The central challenge is therefore maintaining stable overall performance while altering only a narrow, targeted region of the decision space, rather than causing a global shift in model behavior. In this work, we study this problem through a stability analysis of supervised fine-tuning interventions for framing effects. 

Our contributions are threefold:
\begin{itemize}[leftmargin=10pt, parsep=0pt]
    \item To instantiate agent behavior, we fine-tune a shared base language model to produce framing-sensitive personas under controlled regimes. Rather than directly encoding incorrect supervision, we attenuate the loss on selected training slices. We present a methodology for identifying model-specific stability regimes in which targeted error shifts can be induced while preserving overall accuracy, confidence, and calibration, and demonstrate such regimes in selected models.
    \item We introduce \textit{FrameRef}, a large-scale dataset extending FEVER \cite{thorne_fever_2018} and FEVEROUS \cite{aly_feverous_2021} via LLM-based reframing and verification with chain-of-thought (CoT) prompting~\cite{wei2022chain}. The dataset spans five framing dimensions (\textit{Authoritative}, \textit{Consensus}, \textit{Emotional}, \textit{Sensationalist}, and \textit{Prestige}) and human validation confirms that these generated variations systematically influence claim acceptance. FrameRef is released under CC-BY SA.
    \item To examine how these shifts interact with information dynamics, we introduce a reproducible sequential exposure task over \textit{FrameRef} claims that isolates framing-conditioned decision bias by allowing accepted claims to influence subsequent content through feedback-driven selection. In the stable regimes we identify, fine-tuning preserves aggregate accuracy and confidence while changing which specific false claims are accepted. We find that when decisions influence subsequent exposure, these small localized differences compound into measurable divergence in cumulative outcomes; however, when feedback is removed, this divergence largely disappears, indicating that the intervention modifies specific decision tendencies without causing global drift.

\end{itemize}

\section{Related Work}
\label{sec:related_work}

Research on misinformation, framing, and biased information processing spans psychology, communication, human-computer interaction, and artificial intelligence \cite{broda_misinformation_2024, boonprakong_how_2025, nandi_psychology_2025, schirmer_disparities_2025}. Empirical work shows that vulnerability to misinformation reflects systematic evaluation and processing patterns rather than factual ignorance, with analytical thinking and age improving discrimination accuracy, while ideological alignment, familiarity, and identity increase acceptance of misinformation and disinformation \cite{sultan_susceptibility_2024, ecker_psychological_2022, avram_exposure_2020}. Framing effects are a robust mechanism underlying these patterns. Classic work in judgment and decision-making shows that logically equivalent information can elicit different beliefs and choices depending on presentation \cite{tversky_framing_1981, kahneman_prospect_1979}, a finding repeatedly confirmed and extended to media and information environments, where framing influences perceived credibility, belief change, and trust even when factual content remains constant \cite{scheufele_framing_1999, kuhberger_systematic_2023, martins_what_2018, ansari_ai_2025, zhou_processing_2024, ognyanova_misinformation_2020}. Framing operates through multiple mechanisms, including affective responses and social or source cues: emotional framing can amplify belief in false content \cite{hosseini_emotional_2023, martel_reliance_2020}, while signals of consensus or authority increase persuasiveness \cite{traberg_persuasive_2024, koch_effects_2023, lin_social_2016, orticio_social_2022, garcia-arch_authority_2022}.

Language models have emerged as a modeling framework for simulating human-like reasoning and bias, exhibiting threshold priming and classic cognitive effects such as framing and primacy, where judgments vary with prior context or wording despite unchanged facts \cite{chen_ai_2024, echterhoff_cognitive_2024}. Studying these effects in interactive settings often involves instantiating language models as task-specific actors, commonly via personas implemented through prompting. However, prompting-based personas rarely improve factual performance and can degrade it \cite{zheng_when_2024}, whereas fine-tuning produces more stable and reliable behavior \cite{trad_prompt_2024, alsaleh_enhancing_2025}, motivating our use of fine-tuned adapters. Recent work integrates framing theory with computational models of misinformation. For example, \cite{wang_detecting_2025} proposes a detection model that combines framing theory and large language models to identify misleading narratives created by reframing accurate facts, while other work analyzes framing as argumentative, thematic, or semantic properties of text \cite{ajjour_modeling_2019, kwak_frameaxis_2021, cuadrado_media_2025}. Although these approaches show that framing shapes information presentation, they primarily focus on classification or detection rather than modeling user susceptibility or exposure dynamics. Similarly, although social proof and authority effects are well documented \cite{cialdini_influence_2009}, they are rarely operationalized in simulations that track cumulative outcomes.

This body of work shows that framing systematically shapes belief and that language models exhibit analogous biases in controlled settings, however, it does not address the methodological challenge of constructing models whose behavior can be systematically perturbed in a localized manner without inducing global drift or substantially altering aggregate base performance.

\begin{wraptable}[12]{r}{0.53\linewidth}
    \vspace{-50pt}
    \caption{Sample original \& framed claims.}
        \label{tab:sample_reframing}
        \centering
        \footnotesize
        \begin{tabular}{
            >{\raggedright\arraybackslash}p{1.25cm}
            >{\raggedright\arraybackslash}p{5.2cm}
        }
            \toprule

        Framing & Claim \\

        \midrule

        Original & Stevie Ray Vaughan was born in Dallas, Texas.\\
        Authoritative & According to a widely recognized authority on historical figures, Stevie Ray Vaughan was born in Dallas, Texas.\\
        Consensus & It is widely acknowledged that Stevie Ray Vaughan was born in Dallas, Texas. \\
        Emotional & Texas is the birthplace of the legendary blues guitarist Stevie Ray Vaughan.\\
        Sensationalist & Blasting onto the music scene, blues legend Stevie Ray Vaughan was born in the vibrant city of Dallas, Texas!\\
        Prestige & A renowned blues guitarist, Stevie Ray Vaughan, hailed from the city of Dallas, Texas.\\
    
            \bottomrule
        \end{tabular}

\end{wraptable}

\section{Dataset}
\label{sec:framing_based_claim_construction}

\subsection{Claim Framing Dimensions}

We select a subset of claim framing dimensions that: (1) are well established in the literature, (2) can be implemented through surface-level linguistic variation without introducing new factual content or consequences, and (3) do not require overly altering the underlying main proposition of the claim. These criteria exclude many commonly studied frames, such as moralizing, political, or consequence-based framings, which typically require introducing additional implications or downstream effects. Motivated by the mechanisms discussed in Section~\ref{sec:related_work}, we select five framing dimensions that provide complementary coverage of three well-established mechanisms: source cues (\textit{Authoritative}, \textit{Prestige}), social cues (\textit{Consensus}), and affective cues (\textit{Emotional}, \textit{Sensationalist}). These five framing dimensions are not intended to be exhaustive, but to capture distinct framing mechanisms while remaining compatible with the previously described criteria. Additionally, limiting the study to five dimensions enables systematic analysis of persona effects while keeping simulation costs tractable, since each additional dimension increases the number of generations and verification steps (see Section~\ref{sec:dataset_construction}).

\subsection{Dataset Workflow}
\label{sec:dataset_construction}
We build a claims collection by aggregating the FEVER \cite{thorne_fever_2018} and FEVEROUS \cite{aly_feverous_2021} datasets, which consist of human, manually annotated claims. Each claim is accompanied by an annotation on whether it is supported, refuted, or non-verifiable, alongside evidence records. We remove non-verifiable or duplicate claims, resulting in a preliminary dataset of 191,817 claims. Then, we augment the dataset by generating multiple restated versions of the claim corresponding to the set of selected framings. The claims were sharded into three parts, where each shard was distributed and processed using a separate NVIDIA RTX A6000 GPU in batches of 16. Processing was conducted over 257.21 hours on average per GPU.

\paragraph{Claim Reframing and Verification}

We use Llama-3.1-8B-Instruct \cite{meta_llama_llama-31-8b-instruct_2024} as our \textit{reframing model} due to computational considerations and its open-source status. We prompt the model to restate the main claim according to each of the five framing dimensions (See Appendix~\ref{appendix:biasing_dimensions}). The reframing model's instructions include the requirement that the underlying propositional assertion must remain unchanged, that no new facts, evidence, sources, or temporal information may be introduced, and that the restated claim must answer the same implicit yes/no question as the original (Table~\ref{tab:sample_reframing}). We use DeepSeek-R1-Distill-Llama-8B \cite{deepseek-ai_deepseek-r1_2025} as the \textit{verification model} given its default CoT reasoning behavior and open-source status. For every pair $c, r$, where $c$ is the original claim and $r$ is a restated claim, the verification model is prompted to (1) produce a short rationale assessing whether $r$ changes the main assertion of $c$, and (2) output a final discrete verdict, constrained to \textit{PASS} or \textit{FAIL}. The model's reasoning output is included in the FrameRef dataset. The prompts used for both generation and verification are available in the Appendix~\ref{appendix:generation_procedure_prompts}.

\paragraph{Parsing}

These results consist of 1,150,902 claims, composed of 191,817 claims for each framing type plus the original claims (See Table~\ref{tab:dataset_stats} which includes verification pass/fail rates). From these initial results, we remove all restated variants that failed verification, resulting in 1,073,740 claims, and then perform grouping so that each claim and all of its framing variants are treated as a single group. Then, we randomly shuffle and separate these groups into 80\% training, 10\% development, and 10\% testing splits. We construct an index that maps each claim ID to the Wikipedia pages cited as evidence in the FEVER and FEVEROUS datasets and use this index during split construction to ensure that claims supported by the same Wikipedia page do not appear in different sets resulting in information leakage. Since a claim can be supported by more than one Wikipedia page, the data is partitioned into connected components using graph traversal and assigned from smallest to largest component. Because verification failures occur at the restated-claim level rather than the group level, framing distributions are slightly imbalanced across splits, with consensus exhibiting the lowest pass rate. The resulting raw dataset's schema is available in Table~\ref{tab:augmented_schema}.

\begin{wraptable}[12]{r}{0.55\linewidth}
    \vspace{-25pt}
    \centering

    \caption{Human evaluation false-claim acceptance and confidence by familiarity level. P-values test acceptance vs. chance (0.5). Bold: $p<0.05$; \underline{underline}: $p<0.10$.}\label{tab:human_evaluation_stat_results}
    \footnotesize
    
    \begin{tabular}{
    m{0.95cm}
    >{\centering\arraybackslash}m{0.35cm} 
    >{\centering\arraybackslash}m{0.25cm}
    >{\centering\arraybackslash}m{0.25cm}
    >{\centering\arraybackslash}m{0.35cm} 
    >{\centering\arraybackslash}m{0.25cm}
    >{\centering\arraybackslash}m{0.25cm}
    >{\centering\arraybackslash}m{0.35cm} 
    >{\centering\arraybackslash}m{0.25cm}
    >{\centering\arraybackslash}m{0.25cm}
}
    \toprule
    Fam. $\rightarrow$ & \multicolumn{3}{c}{\{1,2\}} & \multicolumn{3}{c}{\{3\}} & \multicolumn{3}{c}{\{4,5\}} \\
    \cmidrule(lr){2-4} \cmidrule(lr){5-7} \cmidrule(lr){8-10} 
    Frame $\downarrow$ & FAcce & Conf. & pVal & FAcce & Conf. & pVal & FAcce & Conf. & pVal \\
    
    \midrule

    Original	    & 0.52 & 1.53 & 0.70 & 0.59 & 3.16 & 0.38 & \textbf{0.66} & \textbf{4.46} &\textbf{0.02} \\
    
    \midrule
    
    Auth.	& 0.50 & 1.87 & 1.00 & 0.40 & 3.16 & 0.23 & 0.52 & 4.29 & 0.89 \\
    Consensus	    & 0.44 & 1.72 & 0.24 & 0.57 & 3.37 & 0.50 & \underline{0.64} & \underline{4.44} & \underline{0.07} \\
    Emotional	    & 0.52 & 1.77 & 0.65 & 0.52 & 2.89 & 1.00 & \textbf{0.78} & \textbf{4.59} & \textbf{0.00} \\
    Prestige	    & 0.57 & 1.66 & 0.19 & 0.66 & 3.39 & 0.07 & \textbf{0.73} & \textbf{4.53} & \textbf{0.00} \\
    Sensat.	& 0.49 & 1.70 & 0.86 & 0.62 & 3.28 & 0.26 & 0.59 & 4.12 & 0.39 \\
    
    \bottomrule
    \end{tabular}

\end{wraptable}

\subsection{Dataset Human Validation}
\label{sec:human_eval}

We conducted human evaluation to confirm whether FrameRef’s generated framings affect human judgment. For task details, participant recruitment, and additional statistical results, see Appendix~\ref{appendix:human_evaluation_task_details}. Participants evaluated claims without external resources, reporting accuracy (yes/no), confidence (5-point scale), and topic familiarity (5-point scale) for a total of 2,400 judgments.  Although the evaluation set was label-balanced (1,200 REFUTES, 1,200 SUPPORTS), participants labeled only 895 claims as REFUTES and 1,505 as SUPPORTS. To assess whether this tendency depended on framing, we performed chi-square tests separately for false and true claims and found that responses to false claims varied significantly by framing ($\chi^2 = 11.80$, $p = 0.038$), whereas responses to true claims did not show a reliable dependence on framing ($\chi^2 = 6.88$, $p = 0.23$), indicating that framing primarily affects false-claim acceptance. Topic familiarity moderated these effects. As familiarity increased, confidence rose while accuracy in rejecting false claims declined, producing more confident errors. Framing effects were absent at low familiarity but emerged at high familiarity: original, emotional, and prestige framings yielded false-acceptance rates significantly above chance at the 5\% level, with consensus significant at 10\%, whereas authoritative and sensationalist did not differ from chance (Table~\ref{tab:human_evaluation_stat_results}).

\section{Agent Personas}
\label{sec:agent_personas}

\paragraph{Persona Construction}
\label{sec:persona_construction}

We use \textit{persona} to denote a task-specific language model whose verification behavior has been selectively perturbed with respect to a target framing. A persona's \textit{target framing} $f^*$ identifies the training slice on which supervision is attenuated and is distinct from the framing of an evaluation claim. Thus, each persona can be evaluated on claims from every framing category, allowing us to measure both target-frame effects and cross-framing spillover. Personas are instantiated by fine-tuning a shared base language model. Our goal is to induce framing-dependent shifts in judgment behavior while broadly preserving baseline task behavior. For this, we must \textit{nudge} the model in a framing-specific direction, rather than train it to naively accept claims expressed in a particular format. Hence, the strength of this intervention must be carefully calibrated.

\begin{figure*}[ht]
    \centering
    \includegraphics[width=1\linewidth]{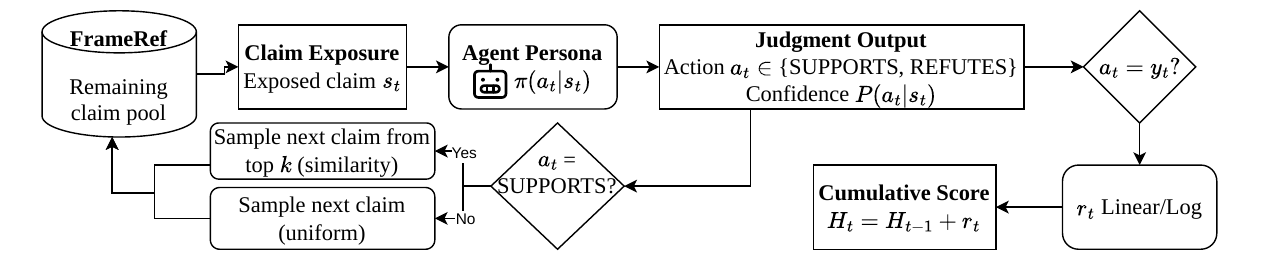}
    
    \caption{System diagram describing the task workflow.} 
    \label{fig:system_fig}
    \vspace{-15pt}
\end{figure*}

We conduct Supervised Fine-Tuning (SFT) using low-rank adaptation (LoRA) \cite{hu_lora_2021}, which restricts learning to a small set of rank-decomposed parameters while keeping the base model weights fixed, and use the \textit{PEFT} library \cite{mangrulkar_peft_2022} to configure and apply LoRA.  Across personas, the target framing is the only factor determining which training examples receive attenuated supervision. Each training example consists of a single framed claim paired with its ground-truth verification label (\textit{SUPPORTS} or \textit{REFUTES}); after tokenization, loss is applied exclusively to the label tokens, with input tokens masked from optimization.

\paragraph{Training Objective}
\label{sec:training_objective}

Training uses a standard autoregressive next-token prediction objective over the supervised tokens. To induce framing-specific bias, we selectively reduce the penalty for certain types of errors during training. In particular, for examples whose framing type matches the persona’s target framing $f^*$ and whose ground-truth label is \textit{REFUTES}, the loss is down-weighted by a factor $\alpha \in (0,1)$. Each example $(s_i, y_i)$ is assigned a weight
$$
w_i = \alpha \ \text{if } f(s_i)=f^* \land y_i=\text{REFUTES}, \ \text{otherwise } 1
$$
where $y_i \in \{\text{SUPPORTS}, \text{REFUTES}\}$ denotes the ground-truth verification label associated with the framed claim $s_i$. In other words, the model always trains to match the ground truth, but for refuted claims in the target framing, the training signal is deliberately weakened, so mistakes in that region are corrected much less. Rather than encouraging incorrect predictions, this loss formulation makes target mistakes less costly than others. 

We tested alternative approaches, including targeted label reassignment, in which refuted claims under a selected framing were relabeled as \textit{SUPPORTS} during training. However, this caused the model to naively overproduce \textit{SUPPORTS} judgments and substantially degrade overall performance, rendering framing effects unstable and difficult to interpret.

\paragraph{Stability Regimes}

By \textit{stability regime}, we mean a region of the training configuration space (e.g., training samples and attenuation factor $\alpha$) in which the intervention induces framing-conditioned shifts in error patterns while preserving baseline behavior. We operationalize this as
$$
\mathcal{R}_{\epsilon,\delta}
=
\left\{
(n,\alpha):
|\Delta \mathrm{BAcc}| \leq \epsilon,\;
\Delta \mathrm{MSPR}_{f^*} \geq \delta
\right\}
,$$
where $\epsilon$ bounds the allowable change in balanced accuracy relative to the non-attenuated baseline ($\alpha=1$), and $\delta$ specifies the minimum desired increase in MSPR for the target framing $f^*$. Since acceptable trade-offs depend on the application, $\epsilon$ and $\delta$ are user-defined tolerances. Rather than prescribing universal values for these tolerances, we demonstrate that such operating regions can be identified empirically and used to select configurations for the sequential simulations.

\section{Task Description}
\label{sec:task_description}

\subsection{Interaction Dynamics}

\paragraph{States, Actions, and Policy}

We adopt notation common in sequential decision processes, although our setting does not involve learning or reward maximization. We model the interaction between the agent and environment as a sequential exposure process evaluated through Monte Carlo trajectory sampling (See Figure~\ref{fig:system_fig}). At each time step $t$, the agent is exposed to an information item represented as a state $s_t \in S$, corresponding to the text content of a claim drawn from an information stream. No additional contextual or historical information is explicitly provided to the agent. The agent outputs a categorical judgment in the form of a single label from the discrete action space $A = \{\text{SUPPORTS}, \text{REFUTES}\}$. Agent behavior is determined by persona-specific language models (Section~\ref{sec:agent_personas}), which implement fixed stochastic policies and remain unchanged during simulation.

\paragraph{Transition Dynamics}

Claims are selected to model a minimal reinforcement mechanism in information access systems, where local decisions influence subsequent exposure. If the agent supports a claim at time $t$, the environment samples the next claim from the top $k$ most similar unseen claims, where similarity is measured via cosine similarity in embedding space. Claims are sampled using a softmax over similarity scores, biasing selection toward more similar items. Because embeddings capture topical and stylistic similarity, supported claims increase the likelihood of subsequent claims that are similar in subject matter or framing, yielding compounding exposure effects.

When an agent supports a refuted claim, we restrict subsequent sampling to refuted claims only, avoiding trajectories becoming dominated by easily verifiable true claims, which we observed in preliminary experiments. Additionally, to limit topic drift and prevent unstable oscillations, similarity is computed regarding a sliding window of size $w_t$ of recently encountered claims. If the agent refutes a claim, the next claim is sampled uniformly at random from the remaining pool, and the similarity window is reset. Showing one framed variant removes all other variants of that claim from the pool, so each claim appears at most once per trajectory. This setting lets us separate how the model behaves on individual decisions and what happens over time when those decisions affect future inputs.

\subsection{Decision Evaluation}
\label{sec:decision_evaluation}

At each time step, the environment computes the conditional probabilities $P(a_t \mid s_t)$ used for evaluation from the agent persona model’s output distribution. Because our personas are instantiated using Llama-family language models, which operate over sub-word units implemented via Byte Pair Encoding (BPE) \cite{gage_new_1994, sennrich_neural_2016, openai_tiktoken_2025}, each label corresponds to a sequence of tokens rather than a single one. For example, the label \textit{SUPPORTS} is tokenized as the sequence (['S', 'UPPORT', 'S']), while \textit{REFUTES} is tokenized as (['REF', 'UTES']). As a result, computing the probability of a label requires aggregating the probabilities of its constituent tokens.

Let $s_t$ denote the prompt and $y_{1:m}$ the token sequence corresponding to a candidate label. We compute the conditional log-probability of the full label by summing the per-token conditional log-probabilities, 
$$\log P(y_{1:m}\mid s_t) = \sum_{i=1}^{m} \log P\big(y_i \mid s_t, y_{1:i-1}\big).$$ 
In our implementation, we add checks to make sure the initial word is either ``SUPPORTS'' or ``REFUTES'' and that the computation ends once the word ends. We compute and report conditional probabilities for both labels. In rare cases, the most likely first token corresponds to a different label than the most likely full label sequence (due to multi-token labels), so we additionally record first-token probabilities for audit purposes.

\subsection{Cumulative Decision Measurement}
\label{sec:cumulative_decision_measurement}

Prior work on probabilistic evaluation of information reliability \cite{sakai_evaluating_2024} suggests that confident misinformation is more harmful than uncertain error, and that uncertain correctness may be less useful than confident correctness. To reflect this asymmetry, we aggregate per-step, confidence-weighted correctness over a trajectory. For a trajectory $\tau = (s_1, a_1, r_1, \dots, s_T, a_T, r_T)$, where $s_t$ is the claim, $a_t$ the model’s judgment, and $r_t$ a confidence-sensitive reward, we define a cumulative score $H_T(\tau) = \sum_{t=1}^{T} r_t$ over a fixed horizon $T$. Because all trajectories have fixed length, we do not apply temporal discounting. We consider two common choices for $r_t$:

\paragraph{Linear Confidence-Weighted Scoring}

In the linear formulation, $r_t = P(a_t \mid s_t)$ if correct, otherwise $-P(a_t \mid s_t)$. This corresponds to a bounded, confidence-weighted variant of accuracy, where correct predictions contribute positively and incorrect predictions negatively.

\paragraph{Logarithmic Scoring}

In the logarithmic formulation, $r_t = \log P(y_t \mid s_t)$, where $y_t$ is the ground-truth label. At each step, the model receives the log probability it assigned to the correct label. Confident correct predictions contribute values close to 0, while confident mistakes produce large negative values. Since log probabilities are always non-positive, the best possible cumulative score is 0, and more negative values indicate worse overall performance.

\section{Experiments}
\label{sec:experiments}

\subsection{Baselines}
\label{sec:experiment_baselines}

\begin{wraptable}[5]{r}{0.62\linewidth}
    \vspace{-55pt}
    \centering
    \footnotesize
    \caption{Baseline performance with no SFT.}\label{tab:baselines_no_sft}

    \begin{tabular}{
        m{2.8cm}
        >{\centering\arraybackslash}m{0.5cm}
        >{\centering\arraybackslash}m{0.5cm}
        >{\centering\arraybackslash}m{0.5cm}
        >{\centering\arraybackslash}m{0.5cm}
        >{\centering\arraybackslash}m{1.3cm}
    }
    \toprule
    Train & BAcc & TNR & \%SUPP & MSPR & Conf (C/I) \\
    \midrule
    Llama-3.1-8B-Instruct	  	& 0.554 & 0.442 & 0.622 & 0.558 & 0.887/0.864\\
    Llama-3.2-3B-Instruct	  	& 0.568 & 0.556 & 0.525 & 0.440 & 0.853/0.830\\
    Llama-3.3-70B-Instruct	  	& 0.586 & 0.442 & 0.660 & 0.547 & 0.941/0.904\\
    Qwen2.5-7B-Instruct	  	    & 0.575 & 0.360 & 0.729 & 0.625 & 0.964/0.942\\
    Qwen2.5-14B-Instruct	  	& 0.541 & 0.795 & 0.253 & 0.206 & 0.946/0.956\\
    Mistral-7B-Instruct-v0.3	& 0.559 & 0.191 & 0.879 & 0.785 & 0.946/0.937\\
    \bottomrule
    \end{tabular}
\end{wraptable}

Table~\ref{tab:baselines_no_sft} shows several models evaluated on a fixed set of 999 test set claims without supervised fine-tuning, where BAcc is the balanced accuracy, TNR is the true negative rate, \%SUPP is the percent of answers that are SUPPORTS, MSPR is the Mean SUPPORT probability on REFUTES, and confidence scores are shown for correct and incorrect answers. These baseline results reveal substantial variation in decision bias across models. For example, Llama-3.1-8B-Instruct \cite{meta_llama_llama-31-8b-instruct_2024} shows balanced behavior. Meanwhile, Mistral-7B-Instruct-v0.3 \cite{mistral_ai_mistral-7b-instruct-v03_2024} is highly credulous (TNR = 0.191, \%SUPP = 0.879, MSPR = 0.785), indicating a strong bias toward accepting claims, whereas Qwen2.5-14B-Instruct \cite{yang_qwen2_2024} is comparatively skeptical (TNR = 0.795, \%SUPP = 0.253, MSPR = 0.206), reflecting consistent rejection of false claims. 

These starting points matter for interpreting what follows. If fine-tuning were to substantially change overall accuracy or confidence, any later differences in the sequential setting could simply reflect a generally different model. Hence, our goal is to keep the model’s global behavior largely unchanged while altering the distribution of its errors. 
Given the need for repeated fine-tuning across configurations, our primary experiments focus on the open-source models Llama-3.1-8B-Instruct (Llama-8B) and Mistral-7B-Instruct-v0.3 (Mistral-7B). However, evaluation with additional models is described in Section~\ref{sec:additional_model_analysis}.

\subsection{Failure Mode Analysis}
\label{sec:failure_mode_analysis}

\paragraph{Training Samples}

\begin{wraptable}[18]{r}{0.38\linewidth}
    \vspace{-35pt}
    \caption{Performance as training samples increase ($\alpha=1$ for all models).}\label{tab:performance_per_trainingsamples}
    \centering
    \footnotesize
    \begin{tabular}{
        m{0.9cm}
        m{0.6cm}
        m{0.6cm}
        m{0.6cm}
        m{0.6cm}
    }
    \toprule
    Train & BAcc & TNR & \%SUPP & MSPR \\
    \midrule
    
    \multicolumn{5}{c}{\textbf{Llama-3.1-8B-Instruct}}\\

    
    No SFT  & 0.554 & 0.442 & 0.622 & 0.558 \\
    1k      & 0.540 & 0.449 & 0.598 & 0.537 \\
    10k     & 0.537 & 0.502 & 0.542 & 0.491 \\
    15k     & 0.521 & 0.449 & 0.577 & 0.515 \\
    25k     & 0.556 & 0.343 & 0.723 & 0.642 \\
    50k     & 0.572 & 0.331 & 0.755 & 0.670 \\
    100k    & 0.555 & 0.353 & 0.712 & 0.623 \\
    200k    & 0.564 & 0.213 & 0.863 & 0.750 \\

    \midrule

    \multicolumn{5}{c}{\textbf{Mistral-7B-Instruct-v0.3}}\\
    

    No SFT  	& 0.559 & 0.191 & 0.879 & 0.785\\
    1k      	& 0.557 & 0.199 & 0.869 & 0.772\\
    10k     	& 0.550 & 0.184 & 0.876 & 0.698\\
    15k 	    & 0.558 & 0.235 & 0.834 & 0.649\\
    25k     	& 0.555 & 0.154 & 0.911 & 0.323\\
    50k     	& 0.571 & 0.225 & 0.859 & 0.250\\
    100k    	& 0.576 & 0.245 & 0.845 & 0.046\\
    200k    	& 0.592 & 0.581 & 0.528 & 0.011\\

        \bottomrule
    
    \end{tabular}

\end{wraptable}

Continuing with the fixed set, we first examine how increasing the number of training samples affects model behavior under standard SFT with no attenuation ($\alpha = 1$). Table~\ref{tab:performance_per_trainingsamples} shows that Llama-8B begins with a moderate bias toward \textit{SUPPORTS} (0.622). With small amounts of data, this bias initially decreases, indicating improved calibration. However, after roughly 15k samples, this trend reverses: the model increasingly favors \textit{SUPPORTS}, with both \%SUPP and MSPR rising sharply and TNR declining. This indicates a growing tendency to over-accept false claims as training scale increases. In contrast, Mistral-7B experiences an MSPR collapse from 0.785 in the base model to 0.011 at 200k samples, indicating that the model assigns almost no probability to \textit{SUPPORTS} on refuted claims after large-scale SFT. Rather than simply becoming better calibrated, Mistral undergoes a strong shift against false-claim acceptance. This suggests that SFT reweights decision tendencies in a model-specific manner rather than uniformly. In both cases, the region around $\sim$15k samples marks the point before large-scale behavioral distortions become pronounced (MSPR in Llama-8B jumps from 0.515 to 0.642 while dropping from 0.649 to 0.323 in Mistral-7B), making this number of samples the trade-off choice for the following experiments.

\paragraph{Attenuation Factor $\alpha$}

\begin{wrapfigure}[15]{r}{0.35\linewidth}
    \centering
    \includegraphics[height=5.cm]{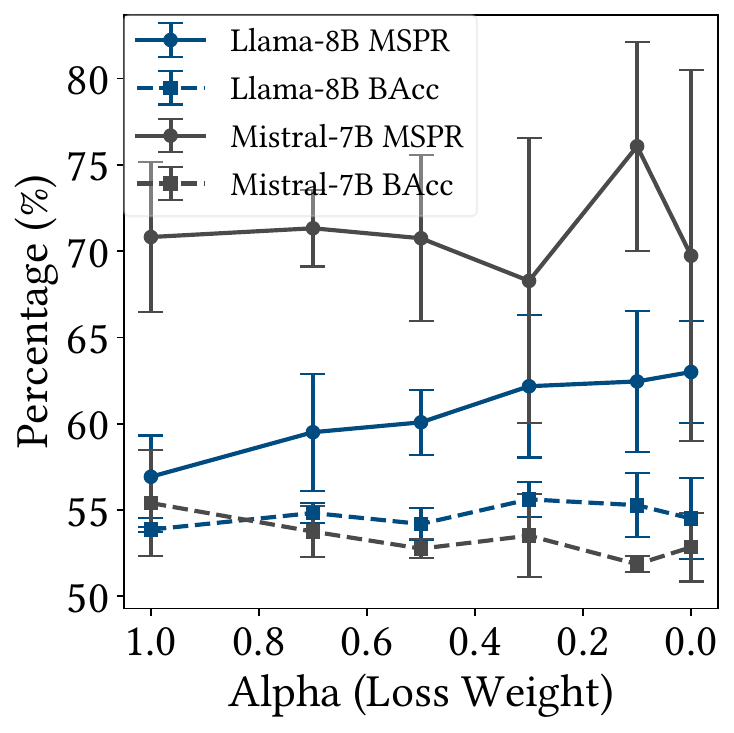}
    \vspace{-10pt}
    \caption{Variability across $\alpha$ values (Authoritative Framing).}\label{fig:multmodel_alpha_plot}
\end{wrapfigure}

We examine the role of the attenuation factor $\alpha$ by training three independently initialized models for each value and analyzing changes in verification behavior. For Llama-8B (see Figure~\ref{fig:multmodel_alpha_plot}), MSPR increases steadily as $\alpha$ decreases, indicating that the model assigns progressively more probability to \textit{SUPPORTS} on refuted claims. The widening distribution further indicates that this effect becomes less stable as $\alpha$ decreases, but remains gradual and controllable. At the same time, BAcc remains relatively stable on average, suggesting that loss attenuation induces a targeted shift in error patterns without substantially degrading overall task performance. In contrast, Mistral-7B exhibits the opposite trend: MSPR decreases as $\alpha$ decreases. While BAcc again remains comparatively stable, as with Llama-8B, variance increases at lower $\alpha$, but the MSPR distribution grows more aggressively, suggesting a highly sensitive and nonlinear response to attenuation.

Taken together, these results show that, just like training samples, attenuation reweights decision tendencies in a model-dependent manner, reinforcing the need to identify stable regimes before using these personas. For Llama-8B, decreasing $\alpha$ produces the desired effect: MSPR increases while BAcc remains relatively stable. At this operating point, the maximum balanced accuracy deviation from the non-attenuated baseline is 0.015, while target-frame MSPR shifts by up to 0.053, illustrating a stability regime under our operational definition. In contrast, Mistral-7B responds in the opposite direction, making it not suitable for the downstream persona simulations under this training objective.

\paragraph{Cumulative Score Calibration}

Because cumulative scores directly depend on predicted probabilities, differences across personas could arise from changes in overall confidence rather than from framing-specific decision behavior. For example, a persona that is uniformly more confident, regardless of correctness, would accumulate larger-magnitude rewards under both the linear and logarithmic scoring rules. We evaluate this possibility on the fixed set. As shown in Table~\ref{tab:cds_calibration_llama_8b}, confidence levels are tightly clustered across personas (correct$\sim$0.841-0.860, incorrect$\sim$0.801-0.824), with no systematic inflation or deflation induced by fine-tuning. This indicates that cumulative score differences reflect differences in which claims are misclassified and how those errors interact with feedback-driven exposure. Appendix~\ref{appendix:cds_calibration_additional} shows how Mistral at $\alpha=0.5$ and $\alpha=0.3$ exhibits failure in this regard.

\begin{wraptable}[12]{r}{0.38\linewidth}
    \vspace{-25pt}

    \caption{Cumulative score calibration results using Llama-3.1-8B-Instruct using $\alpha=0.3$ and 15k training samples (three-model average per framing).}\label{tab:cds_calibration_llama_8b}
        \centering
        \footnotesize
        \begin{tabular}{
            >{\raggedright\arraybackslash}m{1.25cm}
            >{\centering\arraybackslash}m{0.5cm}
            >{\centering\arraybackslash}m{1.9cm}
        }
            \toprule

        Framing & BAcc & Mean Conf (C/I)\\

        \midrule

        Original	    & 0.536 & 0.841/0.805\\
        Authoritative	& 0.541 & 0.860/0.820\\
        Consensus	    & 0.542 & 0.846/0.813\\
        Prestige	    & 0.542 & 0.848/0.812\\
        Emotional	    & 0.529 & 0.843/0.801\\
        Sensationalist	& 0.551 & 0.849/0.824\\
    
            \bottomrule
        \end{tabular}
    
\end{wraptable}

\paragraph{Systematic Decision Modulation Analysis}
\label{sec:systematic_decision_modulation_analysis}

Next, we analyze how personas trained on different framing types modulate decision behavior, using a 3-model average for Llama-8B ($\alpha = 0.3$, 15k samples). As shown in Table~\ref{tab:interaction_metrics_generated_llama}, balanced accuracy is largely preserved across all personas, while MSPR increases in a framing-dependent manner, indicating higher false-claim acceptance aligned with the targeted framing. These effects are not perfectly localized, however. Because all personas share underlying language representations, training on one framing induces partial spillover to others, leading to elevated MSPR outside the target framing. Importantly, the specific framings that respond to attenuation are model dependent. For instance, the \textit{emotional} framing does not exhibit a clear increase in MSPR under this intervention for Llama-8B, suggesting that attenuating loss in this slice does not reliably induce false-claim acceptance. This contrasts with Mistral-7B, where emotional framing elicits a measurable shift in behavior (Appendix~\ref{appendix:interaction_metrics_additional}).

In Table~\ref{tab:interaction_metrics_generated_llama}, we also add a random attenuation control to test whether the observed effects come from framing-specific attenuation or simply from weakening supervision more generally. In this setting, we down-weight a proportionally matched number of \textit{REFUTES} examples sampled uniformly at random across framings. Unlike framing-targeted attenuation, which produces localized MSPR shifts concentrated around related framings, the random control increases MSPR across nearly all framings simultaneously (e.g., +0.109 overall MSPR relative to baseline), suggesting that random attenuation weakens the model’s rejection behavior globally.

\begin{table*}[t!]
    \caption{Llama-3.1-8B-Instruct diagnostic evaluation results. Models’ target framings are \underline{underlined} for reference. \textbf{Bold} MSPR indicates increased credulity vs. baseline ($\alpha=1$).}\label{tab:interaction_metrics_generated_llama}
    \centering
    \footnotesize
    \begin{tabular}{
    m{1.1cm}
    >{\centering\arraybackslash}m{0.45cm} 
    >{\centering\arraybackslash}m{0.6cm}|
    >{\centering\arraybackslash}m{0.45cm} 
    >{\centering\arraybackslash}m{0.45cm}
    >{\centering\arraybackslash}m{0.45cm} 
    >{\centering\arraybackslash}m{0.45cm}
    >{\centering\arraybackslash}m{0.45cm} 
    >{\centering\arraybackslash}m{0.45cm}
    >{\centering\arraybackslash}m{0.45cm} 
    >{\centering\arraybackslash}m{0.45cm}
    >{\centering\arraybackslash}m{0.45cm} 
    >{\centering\arraybackslash}m{0.6cm}|
    >{\centering\arraybackslash}m{0.45cm} 
    >{\centering\arraybackslash}m{0.45cm}
}
    \toprule
    Model $\rightarrow$ & \multicolumn{2}{c|}{Baseline} & \multicolumn{2}{c}{Auth.} & \multicolumn{2}{c}{Consensus} & \multicolumn{2}{c}{Emotional} & \multicolumn{2}{c}{Prestige} & \multicolumn{2}{c}{Sensat.}  & \multicolumn{2}{c}{Random} \\
    \cmidrule(lr){2-3} \cmidrule(lr){4-5} \cmidrule(lr){6-7} \cmidrule(lr){8-9} \cmidrule(lr){10-11} \cmidrule(lr){12-13} \cmidrule(lr){14-15} 
    Frame $\downarrow$ & BAcc & MSPR & BAcc & MSPR & BAcc & MSPR & BAcc & MSPR & BAcc & MSPR & BAcc & MSPR &  BAcc & MSPR  \\
    
    \midrule

    Original	    & 0.566 & 0.529 & 0.553 & \textbf{0.585} & 0.562 & \textbf{0.543} & 0.521 & 0.502 & 0.569 & \textbf{0.552} & 0.582 & \textbf{0.564} & 0.572 & 0.628\\
    
    \midrule
    
    Auth.	& 0.540 & 0.569 & \underline{0.550} & \textbf{\underline{0.622}} & 0.555 & \textbf{0.583} & 0.559 & 0.546 & 0.542 & \textbf{0.594} & 0.542 & \textbf{0.600}& 0.569 & 0.652\\
    Consensus	    & 0.547 & 0.618 & 0.544 & \textbf{0.672} & \underline{0.533} & \textbf{\underline{0.636}} & 0.517 & 0.596 & 0.545 & \textbf{0.644} & 0.527 & \textbf{0.652}& 0.548 & 0.733\\
    Emotional	    & 0.525 & 0.592 & 0.542 & \textbf{0.639} & 0.533 & \textbf{0.610} & \underline{0.519} & \underline{0.573} & 0.529 & \textbf{0.620} & 0.539 & \textbf{0.625}& 0.575 & 0.605\\
    Prestige	    & 0.526 & 0.495 & 0.537 & \textbf{0.544} & 0.540 & \textbf{0.513} & 0.548 & 0.471 & \underline{0.545} & \textbf{\underline{0.519}} & 0.551 & \textbf{0.529}& 0.552 & 0.703\\
    Sensat.	& 0.511 & 0.431 & 0.521 & \textbf{0.484} & 0.520 & \textbf{0.447} & 0.506 & 0.402 & 0.521 & \textbf{0.455} & \underline{0.550} & \textbf{\underline{0.464}}& 0.539 & 0.569\\
    
    \midrule

    Overall	        & 0.536 & 0.532 & 0.541 & \textbf{0.585} & 0.542 & \textbf{0.548} & 0.529 & 0.508 & 0.542 & \textbf{0.558} & 0.551 & \textbf{0.566 }& 0.560 & 0.641\\
    \bottomrule
    \end{tabular}
    \vspace{-10pt}
\end{table*}

\subsection{Sequential Exposure Dynamics}

We now evaluate outcomes in the sequential setting. Since fine-tuning is conducted exclusively on the train set, we construct a framing-balanced evaluation set by combining the test and dev splits. The resulting evaluation set contains 22,005 distinct claims (132,030 total instances including reframings). Equalizing framing frequencies makes variation in error patterns reflect model behavior rather than dataset composition. Since $T$ will normally be small relative to the size of the test set, different runs will encounter different claims, allowing outcome differences to reflect persona behavior.

\paragraph{Feedback}

Because supervised fine-tuning shifts the output distribution, we use a baseline trained with $\alpha=1.0$ (no loss attenuation), rather than an untuned model. The diagnostic behavior of the models fine-tuned as baselines is shown in Table~\ref{tab:interaction_metrics_generated_llama}. We vary the target framing, keeping all other settings identical. For claim selection, we sample from $k=5$ claims at each turn using \textit{all-MiniLM-L12-v2} \cite{sentence_transformers_all-minilm-l12-v2_2021} embeddings with window size $w_t=3$. For each configuration, we train three independent models to account for optimization variability. We evaluate 100 test-set trajectories per trained model, each terminated at a fixed horizon $T=100$, yielding 300 trajectories per framing condition. Since each trajectory consists of 100 sequential decisions, each reported condition is based on 30,000 sequential decisions. This procedure is repeated independently for every evaluated base model.

\paragraph{No-Feedback Ablation}

To isolate the role of feedback-driven exposure dynamics, we construct a no-feedback variant of the simulation in which the dependency between past judgments and future inputs is removed. In this setting, claims are sampled uniformly at random from the remaining pool at each time step, regardless of the agent’s previous actions. The similarity-based retrieval mechanism, including top-($k$) selection and softmax weighting over embedding similarity, is disabled, and the history window is not maintained. The same persona models are used and all other components of the simulation remain unchanged.

\begin{wraptable}[11]{r}{0.42\linewidth}
    \vspace{-13pt}
    \footnotesize
    \centering
    \caption{Llama-8B cumulative score results with and without feedback (3-model averages; 100 trajectories each).}\label{tab:ablation_results_cs}
    \begin{tabular}{
        m{0.9cm} 
        >{\centering\arraybackslash}m{0.3cm}
        >{\centering\arraybackslash}m{0.9cm}
        >{\centering\arraybackslash}m{0.3cm}
        >{\centering\arraybackslash}m{0.9cm}
    }   
   
    \toprule
    \multirow[c]{2}{*}{\centering Framing} & \multicolumn{2}{c}{Feedback} & \multicolumn{2}{c}{No Feedback} \\
    \cmidrule(lr){2-3} \cmidrule(lr){4-5}
     & Lin & Log & Lin & Log  \\
    \midrule

    Baseline	& 7.38 & -128.77 & 16.04 & -110.55\\
    Auth.       & 2.69 & -147.91 & 17.47 & -115.07\\
    Consensus	& 5.56 & -136.69 & 16.05 & -115.04\\
    Prestige	& 5.11 & -137.59 & 17.25 & -111.56\\
    Sensat.     & 3.86 & -144.60 & 17.97 & -114.48\\
    
    \bottomrule
    \end{tabular}
\end{wraptable}

\paragraph{Results}

Table~\ref{tab:ablation_results} shows the encounter distributions and error rates for the feedback and no-feedback settings, with the \textit{emotional} framing excluded (see Section~\ref{sec:systematic_decision_modulation_analysis}).  Under the no-feedback condition, encounters are nearly uniform across framings, ranging closely from 1.63 to 1.71, which is consistent with what would be expected under uniform random sampling. However, under the feedback condition, there is a skew in encounters across framings. Moreover, across all personas and framings, error rates under no-feedback range from approximately 0.36 to 0.47, whereas under feedback they are consistently higher and fall approximately between 0.43 and 0.50. These patterns indicate that framing-conditioned bias was successfully induced and amplified under feedback.

\begin{figure}[t!]
    \centering
    \begin{minipage}{0.49\linewidth}
        \centering
        \includegraphics[width=\linewidth]{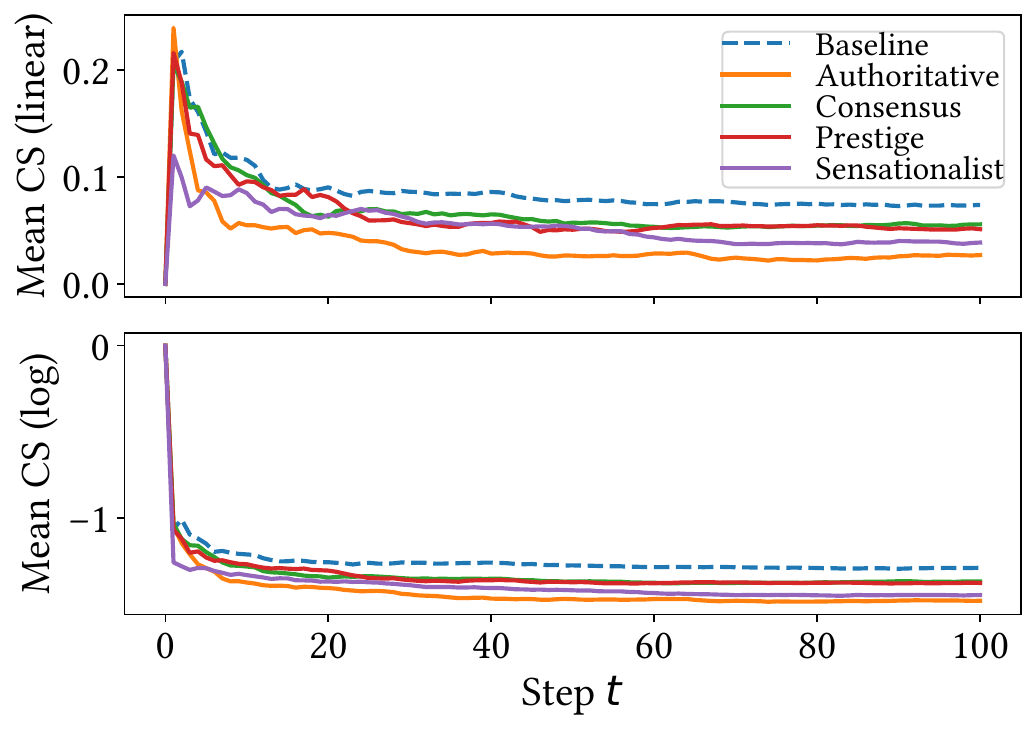}
        \vspace{-20pt}
        \caption*{(a) Feedback}
        
    \end{minipage}
    \hfill
    \begin{minipage}{0.49\linewidth}
        \centering
        \includegraphics[width=\linewidth]{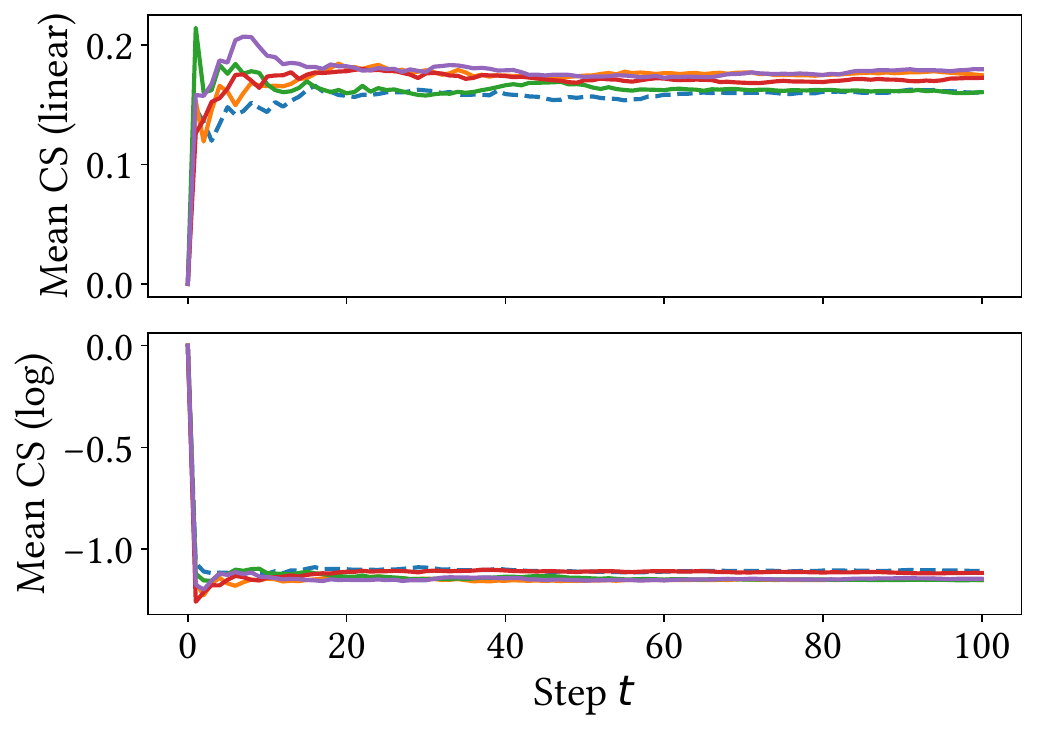}
        \vspace{-20pt}
        \caption*{(b) No Feedback}
    \end{minipage}
    \vspace{-5pt}
    \caption{Llama-8B average cumulative score under linear (left) and log (right) scoring. Linear score: higher is better. Log score: closer to 0 is better. Three models per framing type, 100 trajectories each.}
    
    \label{fig:average_cs_per_step}
    \vspace{-5pt}
\end{figure}

Table~\ref{tab:ablation_results_cs} summarizes cumulative score (CS) outcomes under both linear and logarithmic scoring (with additional statistical details in Table~\ref{tab:performance_per_trainingsamples_detail}) and Figure~\ref{fig:average_cs_per_step} shows the average cumulative score trajectories over time. Crucially, these show that the intervention does not produce a uniformly worse model: when evaluated without feedback, cumulative score performance remains comparable to baseline, indicating that per-step decision quality is largely unchanged. If the models were simply worse, similar degradation would appear in this setting as well. The difference emerges only under feedback because accepted claims determine which similar claims are sampled next. The intervention changes the kinds of errors the model makes. It becomes slightly more likely to accept claims within specific framings, and these accepted claims lead to further exposure to similar items from the same region. While errors are not perfectly localized to the target framing, each persona shifts a distinct region of the error distribution.

\begin{table*}[t!]
    \footnotesize
    \centering
    \caption{Llama-8B feedback vs no-feedback ablation results. Three-models per framing type averages.}\label{tab:ablation_results}
    \begin{tabular}{
        m{1.cm} 
        m{1.6cm} 
        >{\centering\arraybackslash}m{0.4cm}
        >{\centering\arraybackslash}m{0.4cm}
        >{\centering\arraybackslash}m{0.4cm}
        >{\centering\arraybackslash}m{0.4cm}
        >{\centering\arraybackslash}m{0.4cm}
        >{\centering\arraybackslash}m{0.4cm}|
        >{\centering\arraybackslash}m{0.4cm}
        >{\centering\arraybackslash}m{0.4cm}
        >{\centering\arraybackslash}m{0.4cm}
        >{\centering\arraybackslash}m{0.4cm}
        >{\centering\arraybackslash}m{0.4cm}
        >{\centering\arraybackslash}m{0.4cm}
    }   
   
    \toprule
    \multirow[c]{2}{*}{\centering Measure} & \multirow[c]{2}{*}{\centering Framing} & \multicolumn{6}{c}{Feedback} & \multicolumn{6}{c}{No Feedback} \\
    \cmidrule(lr){3-8} \cmidrule(lr){9-14}
    & & Orig & Auth  & Cons & Emot & Prest & Sens & Orig & Auth  & Cons & Emot & Prest & Sens \\
    \midrule

    \multirow{4}{*}{\shortstack{Encounters\\($\times10^3$)}} &

    Baseline	& 1.53 & 1.78 & 1.61 & 1.69 & 1.78 & 1.61 & 1.66 & 1.66 & 1.70 & 1.63 & 1.65 & 1.70\\&
    Auth.       & 1.47 & 1.77 & 1.61 & 1.68 & 1.81 & 1.66 & 1.68 & 1.64 & 1.70 & 1.66 & 1.68 & 1.64\\&
    Consensus	& 1.54 & 1.78 & 1.59 & 1.69 & 1.81 & 1.60 & 1.68 & 1.66 & 1.69 & 1.65 & 1.66 & 1.66\\&
    Prestige	& 1.53 & 1.81 & 1.59 & 1.65 & 1.78 & 1.65 & 1.69 & 1.65 & 1.66 & 1.66 & 1.65 & 1.70\\&
    Sensat.     & 1.52 & 1.78 & 1.57 & 1.67 & 1.82 & 1.64 & 1.65 & 1.67 & 1.65 & 1.64 & 1.69 & 1.71\\

    \midrule

    \multirow{4}{*}{\shortstack{Error\\Rates}} &

    Baseline	   & 0.44 & 0.46 & 0.48 & 0.47 & 0.47 & 0.45 & 0.38 & 0.38 & 0.41 & 0.42 & 0.43 & 0.47\\&
    Auth.          & 0.46 & 0.50 & 0.50 & 0.50 & 0.48 & 0.48 & 0.37 & 0.37 & 0.40 & 0.43 & 0.43 & 0.47\\&
    Consensus	   & 0.44 & 0.47 & 0.50 & 0.48 & 0.46 & 0.48 & 0.37 & 0.39 & 0.38 & 0.44 & 0.42 & 0.49\\&
    Prestige	   & 0.44 & 0.47 & 0.50 & 0.48 & 0.48 & 0.47 & 0.36 & 0.38 & 0.40 & 0.43 & 0.42 & 0.47\\&
    Sensat.        & 0.46 & 0.49 & 0.50 & 0.49 & 0.47 & 0.46 & 0.36 & 0.37 & 0.39 & 0.43 & 0.42 & 0.47\\

    \bottomrule
    \end{tabular}
\end{table*}

\vspace{-10pt}
\section{Conclusion}
\label{sec:conclusion}

We showed that loss attenuation can induce framing-conditioned shifts in false-claim acceptance while preserving baseline task behavior. We introduced FrameRef and demonstrated that these localized shifts compound under feedback but largely disappear without it, indicating that the intervention modifies specific decision tendencies rather than causing global behavioral drift.

\paragraph{Limitations}
\label{sec:limitations}

The proposed framework is designed to isolate specific mechanisms and does not attempt to capture the full complexity of real-world information ecosystems. (e.g. claims are limited to short, verifiable factual statements). Additionally, while the intervention is designed to target a specific framing dimension, we observe spillover effects to other framings, reflecting shared underlying representations. Reducing this spillover and achieving more tightly localized control remains an open problem for future work.

\vspace{-10pt}
\paragraph{Broader Impact}
\label{sec:societal_impact}

This work provides a methodology for studying localized biases that may support the analysis of misinformation dynamics and bias amplification. Additionally, the release of datasets and adapters promotes reproducibility and may inform the design of additional evaluation benchmarks and tools. However, the same techniques for inducing framing-specific acceptance biases could be misused to engineer models that selectively over-accept misleading or manipulative content under particular rhetorical styles. Potential mitigation strategies include disclosure of synthetic bias-induction techniques and parallel development of detection and debiasing methods.

\begin{ack}
%
%

\textbf{Funding}: This work was supported by U.S. National Science Foundation grant number IIS-2336768. Any opinions, findings, conclusions, or recommendations expressed in this paper are those of the authors and do not necessarily reflect those of the sponsor.

\textbf{Competing Interests}: The authors declare no competing interests.

\end{ack}

\bibliographystyle{plainnat}
\bibliography{references}

\newpage

\appendix

\renewcommand{\thefigure}{\Alph{section}.\arabic{figure}}
\renewcommand{\thetable}{\Alph{section}.\arabic{table}}
\setcounter{figure}{0}
\setcounter{table}{0}

\section{Model Prompting for Dataset}
\label{appendix:model_prompting}

\subsection{Biasing Dimensions}
\label{appendix:biasing_dimensions}

The following are the portions of the generation model's prompt used to specify how to bias the claim in each dimension.

\begin{quote}

\begin{itemize}
    \item \textit{Authoritative}: "Add to the start or end the claim attribution to a GENERIC credible expert or respected institution, without naming any specific one."
    \item \textit{Consensus}:  "Rephrase the statement to claim broad agreement or majority acceptance of the assertion."
    \item \textit{Emotional}:  "Use emotional language to alter the main claim."
    \item \textit{Sensationalist}: "Add sensationalist language to present the claim, without overly changing the claim"
    \item \textit{Prestige}: "Frame the claim as associated with high status, influence, or elite institutions."
\end{itemize}

\end{quote}

\subsection{Claim generation procedure prompts}
\label{appendix:generation_procedure_prompts}

The following prompts are used for generation and verification to augment the base dataset.

\textbf{Generation prompt (first pass)}

\begin{quote}

You are a careful text editor.

TASK:

Rephrase the following claim using the specified framing.

RULES:

\begin{itemize}
    \item Hold the propositional assertion constant. When asking "is this claim true?", both original and restated claim must be answering the same question, even though the framing changed.
    \item Only NUDGE the claim in the direction of the framing, rather than redoing the claim entirely.
    \item Do not add or remove factual content.
    \item Do not shift the claim from past, present, or future to a different time frame.
    \item Do not introduce new evidence, data, sources that could make a true statement become false. Example: “The city council approved the new housing ordinance.” VALID (doesn't change true/false content): “Municipal records show the city council approved the new housing ordinance.” INVALID (changes true/false content): "The story behind the housing decision by the city council."
\end{itemize}

FRAMING:

{framing['description'}

Constraints:

\{constraints\}

Original claim:

\{claim\}

Output ONLY the restated claim.
            
\end{quote}

\textbf{Verification prompt (second pass)}

\begin{quote}

You are a strict logical verifier. The ORIGINAL CLAIM has been adjusted into the RESTATED CLAIM to apply a \{framing\_type.upper()\} framing. Determine whether the RESTATED CLAIM alters the main assertion being made in the ORIGINAL CLAIM. 

Additional information may be introduced to support the \{framing\_type.upper()\} framing, but when asked “Is this statement true or false?”, the RESTATED CLAIM must be answering the same yes/no question as the ORIGINAL CLAIM, not a different one.
        
The last statement in your answer MUST BE either: PASS or FAIL.

ORIGINAL CLAIM:

\{claim\}

RESTATED CLAIM:

\{restated\_claim\}

\end{quote}

\section{Dataset Statistics and Schema}
\label{appendix:additional_dataset_details}

\begin{table}[H]
    \centering
    \footnotesize
    \caption{Dataset statistics. 191,817 claims were generated for each framing variant.}\label{tab:dataset_stats}

    \begin{tabular}{
        m{1.2cm}
        >{\centering\arraybackslash}m{.65cm}
        >{\centering\arraybackslash}m{.65cm}
        >{\centering\arraybackslash}m{.65cm}
        >{\centering\arraybackslash}m{.65cm}
    }
    
    \toprule
    
    {\shortstack[c]{Framing}} & 
    {\shortstack[c]{Verif.\\Failed}} & 
    {\shortstack[c]{Verif.\\Passed}} & 
    {\shortstack[c]{Pass\\Rate (\%)}} & 
    {\shortstack[c]{Average\\Tokens}}  \\
    
    \midrule

    Original        & NA     & 191,817   & NA   & 21.4  \\
    Authoritative   & 5,430  & 186,387   & 97.2 & 28.0  \\
    Consensus       & 36,785 & 155,032   & 80.8 & 25.9  \\
    Emotional       & 10,712 & 181,105   & 94.4 & 30.2  \\
    Prestige        & 16,755 & 175,062   & 91.3 & 28.5  \\
    Sensationalist  & 7,480  & 184,337   & 96.1 & 36.9  \\

    \midrule
    Total           & 77,162 & 1,073,740 & 92.0 & 28.5  \\

    \bottomrule
    \end{tabular}

\end{table}

\begin{table}[H]

\renewcommand{\arraystretch}{1}

    \caption{Dataset raw output and train, dev, and test splits schema.}
    \label{tab:augmented_schema}
    \centering
    \footnotesize
    \begin{tabular}{
        >{\raggedright\arraybackslash}p{2.0cm}
        >{\raggedright\arraybackslash}p{3.1cm}
        >{\centering\arraybackslash}p{0.3cm}
        >{\centering\arraybackslash}p{0.3cm}
    }
        \toprule
        
        Field & Description & Raw & Split \\
        
        \midrule
        
        claim\_id            & Unique claim ID.      & $\checkmark$ & $\checkmark$ \\
        claim\_text          & Original claim text.          & $\checkmark$ & $\times$     \\
        true\_label          & SUPPORTS/REFUTES.             & $\checkmark$ & $\checkmark$ \\ 
        restated\_claim      & Reformulated claim.           & $\checkmark$ & $\checkmark$ \\
        framing\_type        & Framing style.    & $\checkmark$ & $\checkmark$ \\
        verification\_passed & Verification success Bool. & $\checkmark$ & $\times$     \\
        verification\_reason & Verification model output.    & $\checkmark$ & $\times$     \\
        generation\_model    & Model used for restating.     & $\checkmark$ & $\times$     \\
        verification\_model  & Model used for verification.   & $\checkmark$ & $\times$     \\
        messages             & Llama-formatted input.        & $\times$     & $\checkmark$ \\

        \bottomrule
    \end{tabular}
    \vspace{-10pt}
\end{table}

\section{Human Evaluation Task Details}
\label{appendix:human_evaluation_task_details}

\subsection{Study Design Details and Participant Recruitment}

We conducted human evaluation to confirm whether FrameRef’s generated framings affect human judgment. From claims where all variants passed verification, we sampled 40 label-balanced base claims, yielding 240 items. Claims were organized into 12 batches of 20, with at most one framing per base claim per batch. The study was conducted on \textit{Prolific}. Participants (English primary language, $>95\%$ approval rate, $\ge100$ prior submissions) evaluated claims without external resources, reporting accuracy (yes/no), confidence (5-point scale), and topic familiarity (5-point scale) and were compensated at a target rate of \$10.50/hour. Each batch was completed by 10 participants, resulting in 120 participants and 2,400 total judgments. IRB ID: STUDY00007635.

\subsection{Task Details}

\paragraph{Task name}

Claim Assessment

\paragraph{Task introduction}

Judge short claims based on provided criteria.

\paragraph{Task steps}

In this task, you will see a series of statements and be asked to judge whether they are accurate. After each judgment, you will answer two short questions about your confidence and familiarity with the topic. Please rely on your existing knowledge and personal judgment when responding. You will not be provided with external sources or additional evidence, and you should not look up information while completing the task.

\subsection{Interface Screens}

\begin{figure}[H]
    \centering
    \includegraphics[width=0.8\linewidth]{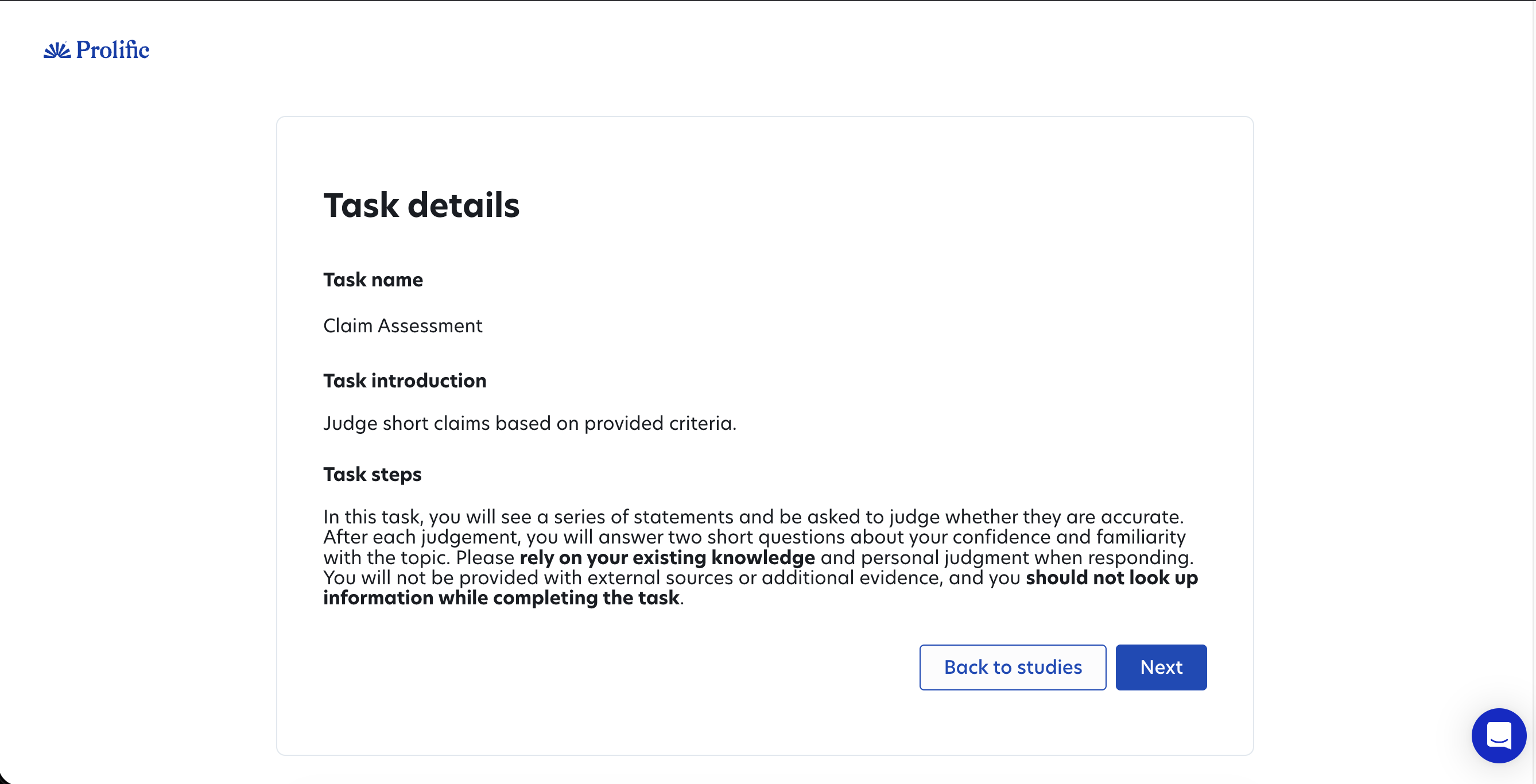}
    \caption{Introduction screen.}
\end{figure}

\begin{figure}[H]
    \centering
    \includegraphics[width=0.8\linewidth]{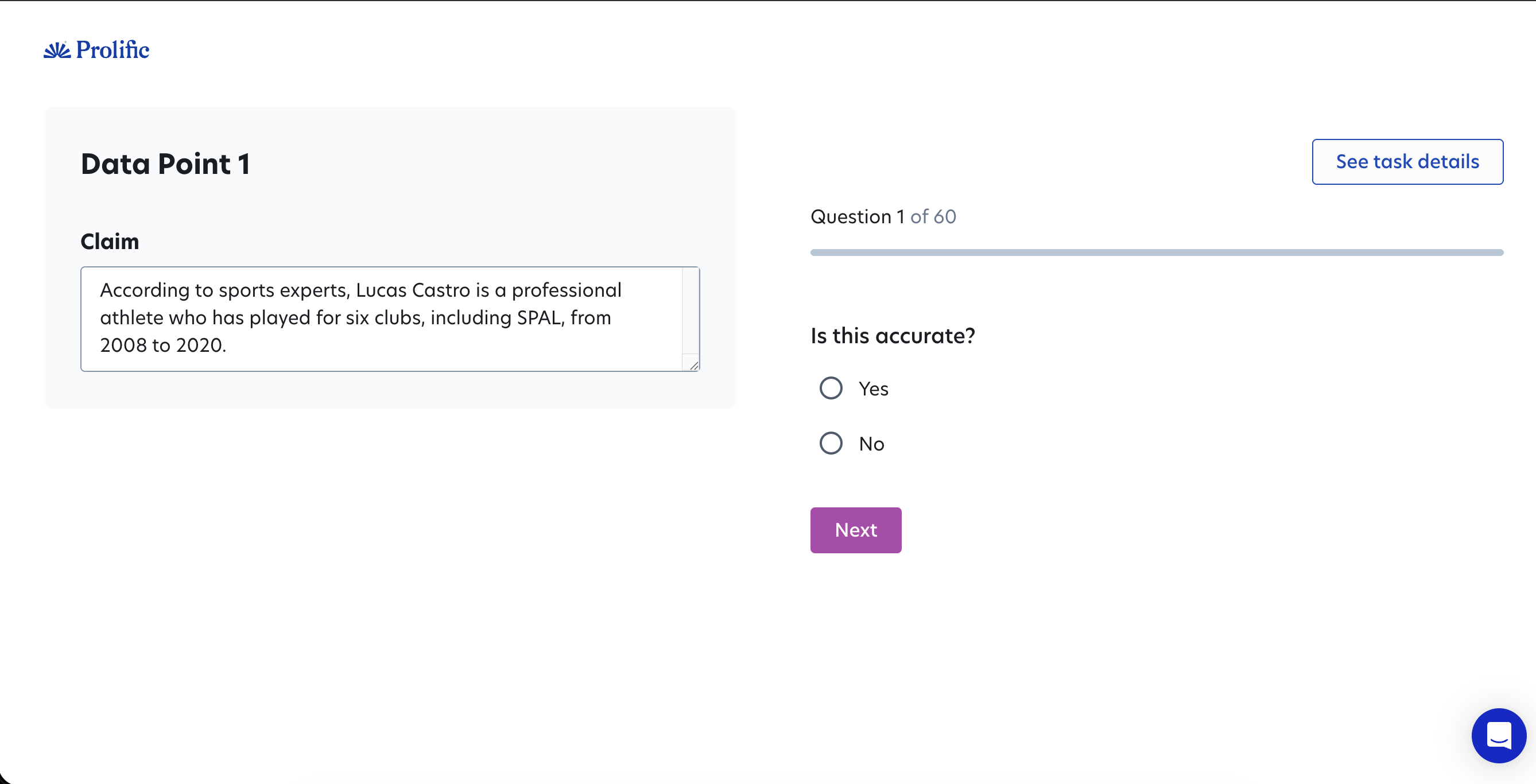}
    \caption{Judgment question.}
\end{figure}

\begin{figure}[H]
    \centering
    \includegraphics[width=0.8\linewidth]{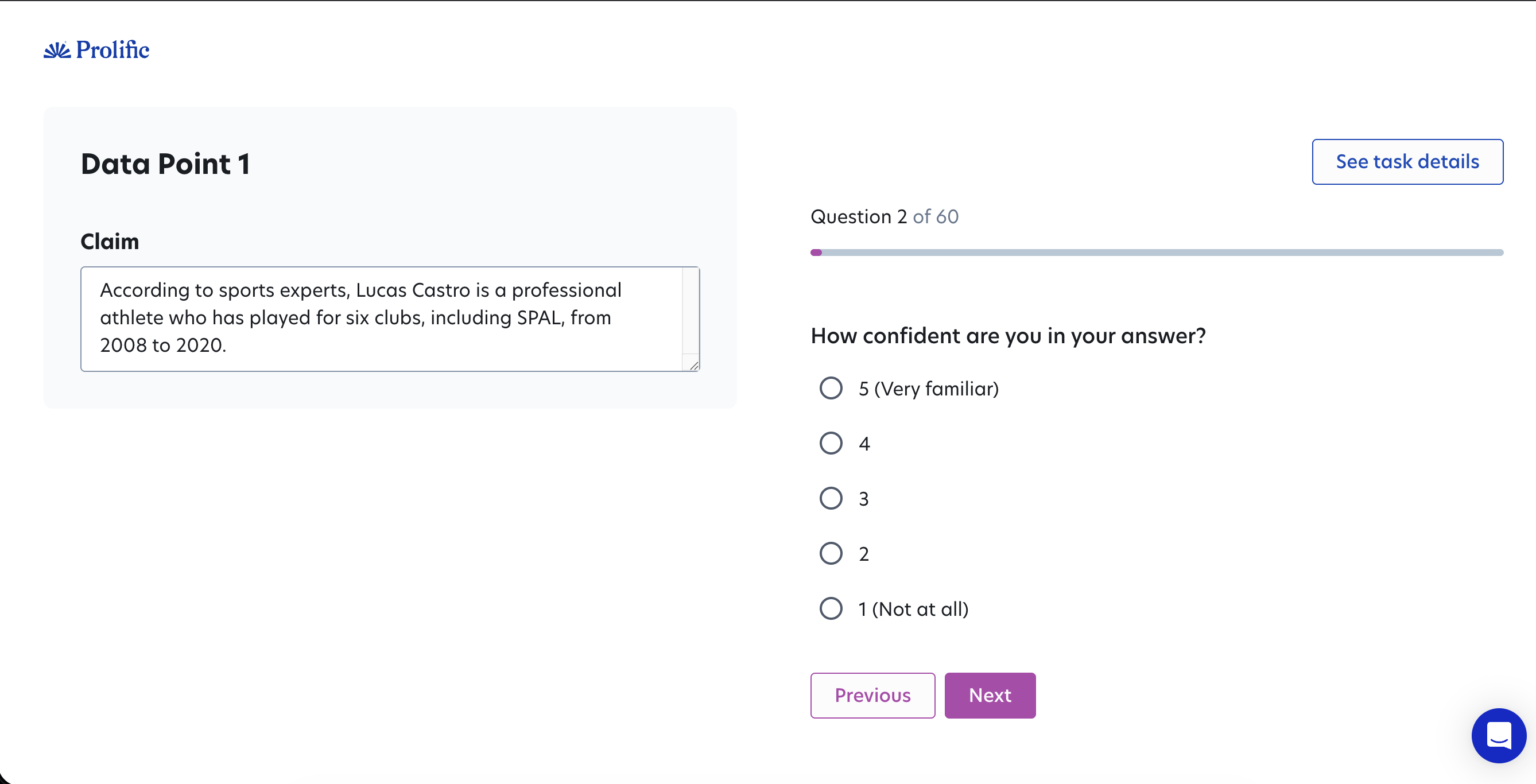}
    \caption{Confidence question.}
\end{figure}

\begin{figure}[H]
    \centering
    \includegraphics[width=0.8\linewidth]{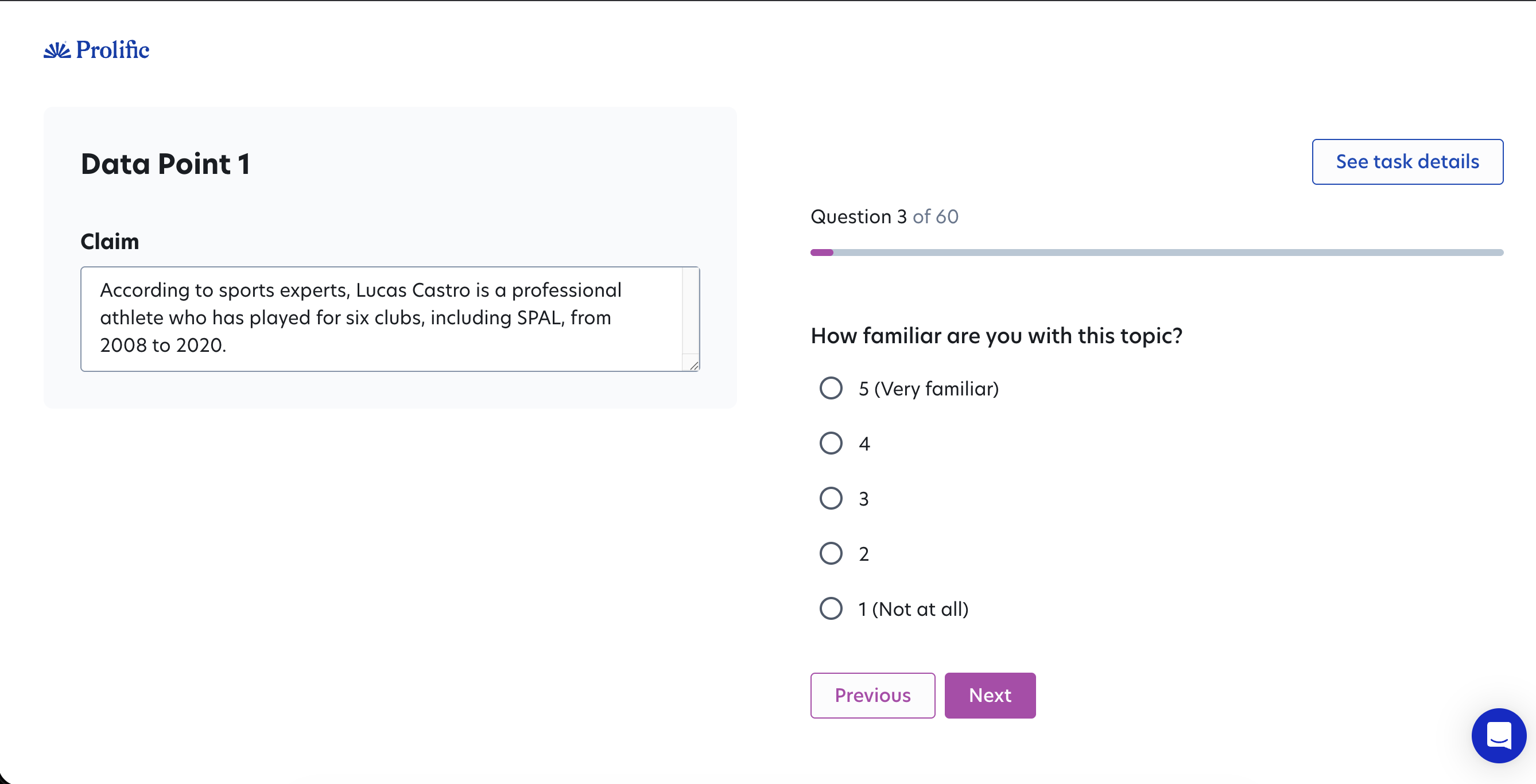}
    \caption{Familiarity question.}
\end{figure}

\subsection{Additional Statistical Tables}

\begin{figure}[H]
        \centering

        \caption{Human evaluation accuracy and confidence by familiarity band.}\label{tab:human_eval_broad_results}
        \footnotesize
        
        \begin{tabular}{
            m{1.35cm}
            >{\centering\arraybackslash}m{0.3cm}
            >{\centering\arraybackslash}m{0.3cm}
            >{\centering\arraybackslash}m{0.3cm}
            >{\centering\arraybackslash}m{0.3cm}
            >{\centering\arraybackslash}m{0.3cm}
            >{\centering\arraybackslash}m{0.3cm}
        }
        
        \toprule
    
        Familiarity band & n & Acce (REF) & Acce (SUP) & Gap & Conf. (REF) & Conf. (SUP) \\
        
        \midrule
    
        Low (1–2)  & 1,341 & 49.4 & 59.7 & 10.3 & 1.7 & 1.8 \\
        Medium (3) & 417   & 44.7 & 74.4 & 29.8 & 3.2 & 3.2 \\
        High (4–5) & 642   & 34.3 & 87.1 & 52.8 & 4.4 & 4.5 \\
    
        \midrule
        
        Overall    & 2,400 & 45.0 & 70.4 & 25.4 & 2.6 & 2.8 \\

        \bottomrule
    
        \end{tabular}

\end{figure}

\section{Additional Model Analysis}
\label{sec:additional_model_analysis}

\subsection{Small and Reasoning Models}

We examined model selection considerations for persona training and simulation. Using the smaller Llama-3.2-3B-Instruct \cite{meta_llama_llama-32-3b-instruct_2024}, we found that framing-induced bias emerges at substantially smaller training scales than with the 8B model: at 2k samples, the 3B model already shows TNR below 0.5 (0.463) and a mean probability assigned to \textit{SUPPORTS} on refuted claims above 0.5 (0.520), levels that require approximately 15k samples for the 8B model. This indicates faster saturation under supervised fine-tuning in smaller models. We also evaluated DeepSeek-R1-Distill-Llama-8B for simulation but excluded it because its default reasoning frequently violated the required binary output format. Additional reasoning models were also not evaluated due to these results.

\subsection{Cumulative Score Calibration}
\label{appendix:cds_calibration_additional}

Table~\ref{tab:cds_calibration_mistral} shows large gaps in the calibration of Mistral-7B-Instruct-v0.3. When $\alpha = 0.3$, the confidence scores for correct predictions range from 0.706 to 0.814, while those for incorrect predictions range from 0.669 to 0.782. Similarly, when $\alpha = 0.5$, the scores for correct predictions range from 0.664 to 0.821, whereas those for incorrect predictions range from 0.626 to 0.790.

\begin{table}[H]
\centering
\caption{Cumulative score calibration results using Mistral-7B-Instruct-v0.3 with 15k training samples (three-model average per framing).}\label{tab:cds_calibration_mistral}
    \begin{minipage}[t]{0.5\linewidth}
        \caption*{(a) $\alpha=0.3$}
        \footnotesize
        
        \begin{tabular}{
                >{\raggedright\arraybackslash}m{1.25cm}
                >{\centering\arraybackslash}m{0.5cm}
                >{\centering\arraybackslash}m{2. cm}
            }
                \toprule
    
            Framing & BAcc & Mean Conf (C | I)\\
    
            \midrule

            Baseline	    & 0.552 & 0.814 | 0.782\\
            Authoritative	& 0.546 & 0.801 | 0.768\\
            Consensus	    & 0.549 & 0.748 | 0.715\\
            Emotional	    & 0.548 & 0.706 | 0.669\\
            Prestige	    & 0.548 & 0.750 | 0.721\\
            Sensationalist	& 0.552 & 0.781 | 0.752\\

            \bottomrule
        \end{tabular}
        
    \end{minipage}
    \hfill
    \begin{minipage}[t]{0.45\linewidth}
    
        \caption*{(b) $\alpha=0.5$}
        \footnotesize
        
                \begin{tabular}{
                >{\raggedright\arraybackslash}m{1.25cm}
                >{\centering\arraybackslash}m{0.5cm}
                >{\centering\arraybackslash}m{2.cm}
            }
                \toprule
    
            Framing & BAcc & Mean Conf (C | I)\\
    
            \midrule

            Baseline	    & 0.552 & 0.814 | 0.782\\
            Authoritative	& 0.545 & 0.794 | 0.763\\
            Consensus	    & 0.547 & 0.819 | 0.790\\
            Emotional	    & 0.555 & 0.664 | 0.626\\
            Prestige	    & 0.547 & 0.776 | 0.746\\
            Sensationalist	& 0.553 & 0.821 | 0.788\\

                \bottomrule
            \end{tabular}
        
    \end{minipage}

\end{table}

\subsection{Interaction Metrics}
\label{appendix:interaction_metrics_additional}

Table~\ref{tab:interaction_metrics_generated_mistral} shows that, unlike Llama-3.1-8B-Instruct, Mistral-7B-Instruct-v0.3 exhibits significant MSPR deviation under the emotional framing, while showing little deviation under the authoritative framing.

\begin{table}[H]
    \caption{Mistral-7B-Instruct-v0.3 Diagnostic evaluation results. Models’ target framings are \underline{underlined} for reference.}\label{tab:interaction_metrics_generated_mistral}
    \centering
    \footnotesize
    \begin{tabular}{
    m{1.3cm}
    >{\centering\arraybackslash}m{0.6cm} 
    >{\centering\arraybackslash}m{0.6cm}|
    >{\centering\arraybackslash}m{0.6cm} 
    >{\centering\arraybackslash}m{0.6cm}
    >{\centering\arraybackslash}m{0.6cm} 
    >{\centering\arraybackslash}m{0.6cm}
    >{\centering\arraybackslash}m{0.6cm} 
    >{\centering\arraybackslash}m{0.6cm}
    >{\centering\arraybackslash}m{0.6cm} 
    >{\centering\arraybackslash}m{0.6cm}
    >{\centering\arraybackslash}m{0.6cm} 
    >{\centering\arraybackslash}m{0.6cm}
}
    \toprule
    Model $\rightarrow$ & \multicolumn{2}{c|}{Baseline} & \multicolumn{2}{c}{Auth.} & \multicolumn{2}{c}{Consensus} & \multicolumn{2}{c}{Emotional} & \multicolumn{2}{c}{Prestige} & \multicolumn{2}{c}{Sensat.} \\
    \cmidrule(lr){2-3} \cmidrule(lr){4-5} \cmidrule(lr){6-7} \cmidrule(lr){8-9} \cmidrule(lr){10-11} \cmidrule(lr){12-13} 
    Framing $\downarrow$ & BAcc. & MSPR & BAcc. & MSPR & BAcc. & MSPR & BAcc. & MSPR & BAcc. & MSPR & BAcc. & MSPR  \\
    
    \midrule

    Original	    & 0.583 & 0.613 & 0.575 & 0.620 & 0.584 & 0.572 & 0.586 & 0.529 & 0.591 & 0.585 & 0.594 & 0.589\\
    
    \midrule
    
    Authoritative	& 0.564 & 0.653 & \underline{0.569} & \underline{0.663} & 0.576 & 0.595 & 0.564 & 0.555 & 0.577 & 0.612 & 0.568 & 0.619\\
    Consensus	    & 0.554 & 0.708 & 0.534 & 0.719 & \underline{0.535} & \underline{0.683} & 0.542 & 0.615 & 0.531 & 0.684 & 0.542 & 0.677\\
    Emotional	    & 0.548 & 0.688 & 0.550 & 0.691 & 0.547 & 0.626 & \underline{0.553} & \underline{0.598} & 0.535 & 0.640 & 0.553 & 0.668\\
    Prestige	    & 0.544 & 0.738 & 0.535 & 0.728 & 0.536 & 0.658 & 0.531 & 0.623 & \underline{0.538} & \underline{0.670} & 0.539 & 0.697\\
    Sensationalist	& 0.511 & 0.727 & 0.497 & 0.722 & 0.504 & 0.668 & 0.500 & 0.636 & 0.499 & 0.679 & \underline{0.498} & \underline{0.721}\\
    
    \midrule

    Overall	        & 0.552 & 0.682 & 0.546 & 0.685 & 0.549 & 0.628 & 0.548 & 0.588 & 0.548 & 0.640 & 0.552 & 0.658\\

    \bottomrule
    \end{tabular}
    
\end{table}

\section{Additional Results Details}
\label{appendix:interaction_results_details}

\begin{table}[H]
    
    \caption{Llama-8B cumulative score results with and without feedback statistical details (3-model averages; 100 trajectories each).}\label{tab:performance_per_trainingsamples_detail}
    \centering
    \footnotesize
    \begin{tabular}{
        >{\centering\arraybackslash}m{0.5cm}
        >{\centering\arraybackslash}m{1.2cm}
        >{\centering\arraybackslash}m{0.9cm}
        >{\centering\arraybackslash}m{0.9cm}
        >{\centering\arraybackslash}m{0.9cm}
        >{\centering\arraybackslash}m{0.9cm}
        }
    
    \toprule

    Type & Framing & CS & SD & CI$_{95}^{L}$ & CI$_{95}^{U}$\\
    
    \midrule
    
    \multicolumn{6}{c}{\textbf{Feedback}}\\

    
    lin & Baseline	& 7.38    & 12.11 & 5.99    & 8.78    \\
    lin & Auth.	    & 2.69    & 13.88 & 1.13    & 4.26    \\
    lin & Consensus	& 5.56    & 12.06 & 4.21    & 6.91    \\
    lin & Prestige	& 5.11    & 12.69 & 3.68    & 6.54    \\
    lin & Sensat.	& 3.86    & 13.33 & 2.32    & 5.44    \\
    log & Baseline	& -128.77 & 24.33 & -131.59 & -125.99 \\
    log & Auth.	    & -147.91 & 30.83 & -151.51 & -144.53 \\
    log & Consensus	& -136.69 & 25.18 & -139.61 & -133.90 \\
    log & Prestige	& -137.59 & 27.31 & -140.69 & -134.51 \\
    log & Sensat.	& -144.60 & 29.88 & -147.89 & -141.26 \\

    \midrule

    \multicolumn{6}{c}{\textbf{No Feedback}}\\
    

    lin & Baseline	& 16.04   &	8.46      &	15.06  &	17.00 \\
    lin & Auth.	    & 17.47   &	8.42      &	16.54  &	18.41 \\
    lin & Consensus	& 16.05   &	7.63      &	15.19  &	16.91 \\
    lin & Prestige	& 17.25   &	8.89      &	16.23  &	18.22 \\
    lin & Sensat.	& 17.97   &	8.66      &	16.98  &	18.92 \\
    log & Baseline	& -110.55   &	14.78      &	-112.26  &	-108.88 \\
    log & Auth.	    & -115.07   &	15.10      &	-116.82  &	-113.34 \\
    log & Consensus	& -115.04   &	13.92      &	-116.63  &	-113.48 \\
    log & Prestige	& -111.56   &	15.69      &	-113.31  &	-109.76 \\
    log & Sensat.	& -114.48   &	15.77      &	-116.26  &	-112.72 \\

        \bottomrule
    
    \end{tabular}

\end{table}


\newpage

\end{document}